# Optimal load sharing in bioinspired fibrillar adhesives: Asymptotic solution

Harman Khungura and Mattia Bacca[*]

*Mechanical Engineering Department, University of British Columbia, Vancouver BC V6T1Z4, Canada*

**Abstract**

We propose here an asymptotic solution defining the optimal compliance distribution for a fibrillar adhesive to obtain maximum theoretical strength. This condition corresponds to that of equal load sharing (ELS) among fibrils, *i.e.* all the fibrils are carrying the same load at detachment, hence they all detach simultaneously. We model the array of fibrils as a continuum of linear elastic material that cannot laterally transmit load (analogous to a Winkler soil). Ultimately, we obtain the continuum distribution of fibril's compliance in closed-form solution and compare it with previously obtained data for a discrete model for fibrillar adhesives. The results show improving accuracy for an incremental number of fibrils and smaller center-to-center spacing. Surprisingly, the approximation introduced by the asymptotic model show reduced sensitivity of the adhesive strength with respect to misalignment and improved adhesive strength for large misalignment angles.

**Introduction**

The presence of fibrillar structures has long been noted in bio-inspired dry adhesion. Throughout nature, various species of insects and larger animals such as geckos use this reversible and repeatable adhesion to scale vertical surfaces [1-4]. Research into the gecko toe pad has revealed hierarchical fibrillar structures comprised of lamellae, setae and spatulae. Spatulae are the end structures of this system and are sub-micrometers in diameter while still being several micrometers in length [5]. There is evidence to suggest that van der Waals interactions play a strong role in the adhesion of these fibrillar structures [6-7]. Surface chemistry is not considered to be a significant contributor to this adhesion, since van der Waals forces are weak and short-ranged. This suggests the importance of mechanics in determining the strength of these adhesives.

Investigations across different sized animals and fibrillar structures have revealed a strong inverse scaling effect, where the adhesive force increases with a greater subdivision of terminal structures [8-9]. This is due to large terminal structures having a smaller surface area to volume ratio and being greatly sensitive to variations in tip geometry. Despite intimate contact, these variations can induce stress concentrations, which in turn reduce the detachment force since failure propagates from an interfacial defect. This is not the case with smaller structures as the interfacial stress distribution becomes more uniform [9-10]. This phenomenon leads to flaw insensitivity and local detachment only occurs when the theoretical adhesive strength is exceeded. Flaw insensitivity is then the condition at which stress concentration does not occur, condition also called equal load sharing (ELS), given the uniform distribution of interfacial stress.

---

[*] Corresponding author.
 E-mail address: mbacca@mech.ubc.ca

Besides the scaling argument mentioned above, another design strategy incorporated into dry adhesive engineering prototypes is to render the interfacial stress uniform by adoption of soft fibrils tips, while keeping their stalks sufficiently rigid to avoid their mutual adhesion (fibril condensation). This strategy basically relies on a stiffness gradient along the longitudinal axis of the fibrils. There is evidence to suggest that such a stiffness gradient exists throughout the setae of the adhesive pads of insects, which soft tips and a compliant backing layer have the additional ability to better conform to the surface roughness [11-13]. A similar approach has also been investigated for synthetic mimics to define new engineering design principles for these adhesives to enter a flaw-tolerant regime [14-15].

All the above-mentioned investigations have been mainly focused on improving the design of the single adhesive unit. Less attention has been dedicated to the design of these adhesives at the array scale, where the mutual interaction of fibrils has shown to significantly affect the global performance of the adhesive [13, 16]. The use of stiffness grading at the array scale, to achieve ELS, was proposed by Bacca and coworkers [16], suggesting a scenario where the load carried by each fibril is uniformly distributed throughout the array, hence reaching the maximum load bearing capability of the adhesive with the simultaneous detachment of all the fibrils. They proposed a numerical model to determine the optimal compliance distribution of the fibrils within the array. This, however, comes with a computational cost and without the numerous benefits of an analytical solution. Of particular importance is the link between the physical properties of the adhesive and the contrast between the softest fibrils and the stiffest ones. We propose in this paper an asymptotic model to overcome these limitations, providing an analytical solution for the optimal compliance distribution for reversible adhesives in normal loading. In this model, we represent the collection of fibrils as a continuum made of a linear elastic material that cannot laterally transmit load, analogous to a Winkler soil. The solution can be obtained in closed form for various shapes of the contact region between the adhesive and the adhered substrate (from now on called "array shape"). We derive this solution for the case of a circular array and for that of a square array and finally demonstrate its validity in providing ELS for both cases.

The results obtained can be generalized to the case of an adhesive interface populated by brittle bonds of varying compliance. When the distribution of compliance matches the one proposed from our analytical model, the interface will reach maximum theoretical strength by minimization of stress concentration at its surface. This phenomenon has been experimentally observed in shear adhesion [17].

**Theoretical model for fibrillar adhesives**

We utilize here the model developed by Bacca *et al.* [16]. We consider a fibrillar adhesive to be composed of a backing layer (BL), described as a linearly elastic half-space, and an array of $N$ linearly elastic fibrils protruding from its surface, as sketched in Figure 1. The tip of each fibril is in contact with a rigid surface (RS), constituting the adhered surface, and the adhesive force $f_i$ is that transmitted from the RS to the BL through fibril $i$. The total adhesive force of the interface is then

$$F = \sum_{i=1}^{N} f_i \tag{1}$$

We simulate the mechanism of detachment of the adhesive interface considering displacement control, *i.e.* by prescribing the separation distance between the BL and the RS and calculating then the force $F$ required to achieve such a separation. In the presence of misalignment along the $x$ axis, as indicated in Figure 1, the displacement prescribed at the tip of each fibril is

$$u_i = \bar{u} + \tan\theta \, x_i \tag{2}$$

with $\bar{u}$ the prescribed separation of the RS from the adhesive, $\theta$ the misalignment angle and $x_i$ the position of fibril $i$ with respect to the line of minimum displacement, respectively. Assuming linear elastic response from the materials involved, the displacement observed at the tip of fibril $i$ can be described as (see Bacca *et al.* [16] for more details)

$$u_i = C_{ij} f_j \tag{3a}$$

(repeated indexes indicate a sum, unless otherwise specified) with

$$C_{ij} = \frac{16}{3\pi^2 E^* a_i} + c_i, \quad i = j \tag{3b}$$

$$C_{ij} = \frac{1}{\pi E^* r_{ij}}, \quad i \neq j \tag{3c}$$

and $f_j$ the force applied to a generic fibril $j$. In this equations, $E^* = E/(1-v^2)$, with $E$ and $v$ the Young modulus and the Poisson ratio of the material composing the BL, respectively. $a_i$ is the radius of the stalk of fibril $i$, while $c_i = h_i/(\pi a_i^2 E_{f,i})$ is the compliance of the same fibril with $h_i$ the length of its stalk, $E_{f,i}$ the Young modulus of the material composing it and $r_{ij}$ the distance between fibrils $i$ and $j$. Reverting then Eq. (3), with $\boldsymbol{K} = \boldsymbol{C}^{-1}$, we have

$$f_i = K_{ij} u_j \tag{4}$$

Prescribing a separation $\bar{u}$, we can then substitute the displacements calculated from Eq. (2) into Eq. (4) and the results of this into Eq. (1) to finally obtain the required force $F$ applied to the adhesive. For the case of an array of identical fibrils, the adhesive force indicated in Eq. (4) computes to a different value for every fibril. The fibrils at the periphery of the array experience a higher load compared to the ones at the center of the array. As indicated in Figure 1, once the force $f_i$, transmitted by fibril $i$, reaches its maximum, $f_{i,max}$, the fibril will detach. For the example of identical fibrils, we can observe the peripheral fibrils detaching prior to those at the center of the array [16].

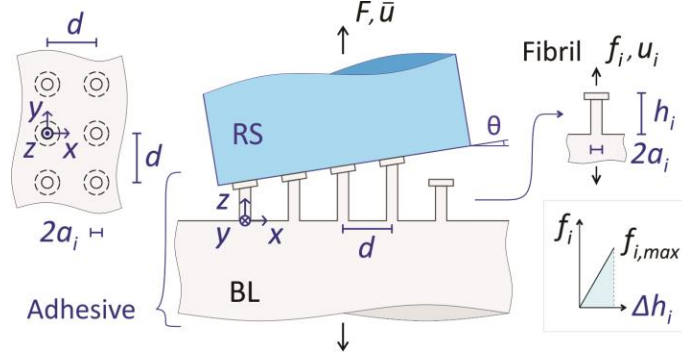

**Figure 1** Sketch of the model adopted. The adhesive adheres to a rigid surface (RS) and is composed of a backing layer (BL) and an array of cylindrical fibrils.

**Equal load sharing**

As theorized by Bacca *et al.* [16], an array of non-identical fibrils can achieve equal load sharing (ELS) if the fibrils have properties tailored to a specific criterion. The condition of ELS corresponds to the situation where the adhesive interface achieves its maximum theoretical strength of

$$F_{max} = \sum_{i=1}^{N} f_{i,max} \tag{5}$$

This implies every fibril reaches its maximum detachment force at the same time. This condition is satisfied when the compliance of every fibril is equivalent to its optimal value, hence $c_i = c_i^*$ $\forall i$, with $c_i^*$ the optimal compliance of fibril $i$. Let us then assume all fibrils have identical tips, hence $f_{i,max} = f_{max}$ $\forall i$, and mean fibril compliance as

$$c_m = \frac{h}{\pi E_f a^2} \tag{6}$$

with $h$ and $a$ the mean stalk length and stalk radius of the fibrils and $E_f$ a nominal value for the Young modulus of the fibrils. As demonstrated in *Appendix A*, we can formulate the optimal fibril compliance distribution, assuming a design misalignment $\theta_d$ as

$$\frac{c_i^*}{c_m} = 1 + \rho \frac{R}{h} \frac{E_f}{E^*} (\psi_m - \psi_i) + \tan\theta_d \left(\frac{x_i - x_m}{\Delta h_{max}}\right) \tag{7a}$$

where

$$\rho = \frac{\pi a^2}{d^2} \tag{7b}$$

is the fibril (stalk) density, with $d$ the center-to-center spacing of the fibrils (Figure 1),

$$\psi_i = \frac{\Gamma}{N} \left(\frac{16}{3\pi} \frac{R}{a_i} + \sum_{j \neq i} \frac{R}{r_{ij}}\right) \tag{7c}$$

is a factor that depends on the position of the fibril in the array and which mean value is $\psi_m$,

$$\Gamma = \frac{Nd^2}{\pi R^2} \tag{7d}$$

is a geometrical factor associated with the array shape, with $R$ the radius of the circle included in the contact region,

$$\Delta h_{max} = c_m f_{max} \qquad (7e)$$

is the average fibril elongation at detachment, and $x_m = \frac{1}{N}\sum_i x_i$ is the average fibril position along $x$ axis. For a dense array with a large number of fibrils, we can assume $A \simeq Nd^2$, giving $\Gamma \simeq 1$ for a circular array and $\Gamma \simeq 4/\pi$ for a square array. Also, under the same hypothesis, $x_m$ becomes the $x$ coordinate of the centroid of the area covered by the fibrillar array.

The solution proposed in Eq. (7) is valid for any array shape. However, this method comes with the computational cost of roughly $N^2$ operations that must be performed for any given adhesive. In the coming sections, we propose an asymptotic method to arrive at an analytical solution for the optimal compliance distribution of any sized array and that can be calculated for any arbitrary array shape.

**Asymptotic analysis**

An array of fibrils can be considered asymptotically as a continuum of linearly elastic material without lateral load transmission, analogous to a Winkler soil. The contact stress $\sigma$ transmitted from the RS to the BL through van der Waals adhesion is defined as a function of the position, $x$ and $y$ within the region of contact, via the function $\sigma(x,y)$. This stress is homogenized from the load applied to the single fibril $i$ via the relation $\sigma(x_i, y_i) = f_i/d^2$. It can be deduced that the accuracy of this asymptotic homogenization is inversely proportional to $d$.

Let us assume now again ELS condition at complete detachment, with all fibrils exerting identical force $f_i = f_{max}\ \forall i$. In this case we have homogenous contact stress, at detachment calculated as

$$\sigma(x,y) = f_{max}/d^2 \qquad (8)$$

at any point $(x,y)$ in the contact region. Consider then the function $c^*(x,y)$ describing the optimal distribution of fibril compliance, with $c^*(x_i, y_i) = c_i^*$. As explained in *Appendix B*, for a design misalignment $\theta_d$, this function can be obtained in its dimensionless form as

$$\frac{c^*(x,y)}{c_m} = 1 + \rho \frac{R}{h}\frac{E_f}{E^*}[\tilde{u}_{BL,m} - \tilde{u}_{BL}(x,y)] + \tan\theta_d \left(\frac{x-x_m}{\Delta h_{max}}\right) \qquad (9a)$$

where

$$\tilde{u}_{BL}(x,y) = \frac{E^* d^2}{f_{max} R} u_{BL}(x,y) \qquad (9b)$$

is a dimensionless form of the displacement of the BL, $u_{BL}(x,y)$, and which average value over the contact region is $\tilde{u}_{BL,m}$. $u_{BL}(x,y)$ is the displacement of a linearly elastic half-space subject to the uniform stress at Eq. (8) applied over the contact region.

Comparing the asymptotic model at Eq. (9) with that in Eq. (7), we can deduce that

$$\psi_i = \tilde{u}_{BL}(x_i, y_i) \qquad (10)$$

By incrementing the number of fibrils per unit area and assuming small variation in stalk radius among fibrils ($a_i \simeq a$), the results obtained with Eq. (7) approach those obtained with Eq. (9), asymptotically. This is demonstrated in *Appendix C*.

## Results

In this section we will obtain the solution of Eq. (9) for the case of a fibrillar adhesive with circular array and for the case of a rectangular array. Since the misalignment in an application is undeterminable, in magnitude and sign, we assume $\theta_d = 0$. This will guarantee ELS for $\theta = 0$ and improved load sharing for $\theta \neq 0$ [16]. Later, we will explore how the adhesive strength reduces for increasing $\theta$.

*Circular array of fibrils*

For the case of a circular array, the function expressed in Eq. (9b) becomes [18]

$$\tilde{u}_{BL}(r) = \frac{4}{\pi} \breve{E}\left(\frac{r}{R}\right) \tag{11}$$

where $x$ and $y$ was replaced by $r$, thanks to radial symmetry. In Eq. (11), $\breve{E}(r/R) = \int_0^{\pi/2}(1 - r^2\sin^2(\theta)/R^2)^{1/2}d\theta$ is the complete elliptic integral of the second kind. Eq. (9) then becomes,

$$\frac{c^*(\tilde{r})}{c_m} = 1 + \rho \frac{E_f}{E^*}\frac{4R}{\pi h}\left[\breve{E}_m - \breve{E}(\tilde{r})\right] \tag{12}$$

with $\breve{E}_m = 2\int_0^1 \breve{E}(\tilde{r})\tilde{r}d\tilde{r}$, and $\tilde{r} = r/R$.

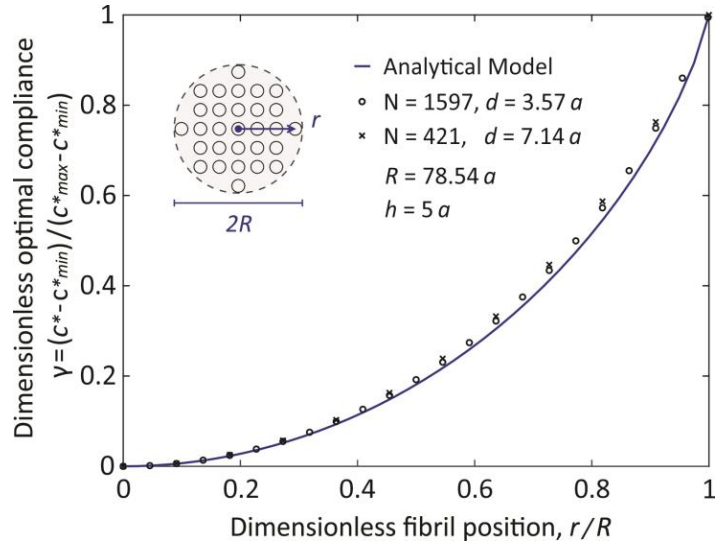

**Figure 2** Optimal compliance distribution for fibrils in a circular array of radius $R$. Crosses and circles indicate the solution obtained with the numerical method, from Eq. (7), while the solid line reports the results from the proposed analytical model, from Eq. (12).

The optimal compliance described in Eq. (7) and (9) have generally maximum at the perimeter of the contact region and minimum at the center. Identifying the former with $c^*_{max}$ and the latter with $c^*_{min}$, we report the geometrical distribution of fibril compliance in dimensionless form with

$$\gamma = \frac{c^* - c^*_{min}}{c^*_{max} - c^*_{min}} \tag{13}$$

in Figure 2. In this figure, the solid line indicates the results obtained from the proposed asymptotic solution at Eq. (12), while the symbols indicate the results obtained from Eq. (7). We assumed $E_f = E$, $\nu = 0.5$, $R = 78.54\,a$ and $h = 5\,a$. For the numerical analysis, we considered the case of $d = 3.57\,a$ and $N = 1597$, giving $\rho = 0.25$ from Eq. (7b), and that of $d = 7.14\,a$ and $N = 421$, with $\rho = 0.06$. We can observe an increasing accuracy as the fibril density $\rho$ increases.

*Rectangular array of fibrils*

For a rectangular array of size $2L$ by $2l$, with aspect ratio $\eta = l/L$ ($\eta \geq 1$) and $L = R$, Eq. (12) becomes [18]

$$\tilde{u}_{BL}(x,y) = \frac{1}{\pi} \Phi\left(\frac{x}{L}, \frac{y}{l}\right) \tag{14a}$$

with

$$\Phi(\tilde{x}, \tilde{y}) = (\tilde{x}+1) \ln\left[\frac{\eta(\tilde{y}+1)+\sqrt{\eta^2(\tilde{y}+1)^2+(\tilde{x}+1)^2}}{\eta(\tilde{y}-1)+\sqrt{\eta^2(\tilde{y}-1)^2+(\tilde{x}+1)^2}}\right] + \eta(\tilde{y}+1) \ln\left[\frac{(\tilde{x}+1)+\sqrt{\eta^2(\tilde{y}+1)^2+(\tilde{x}+1)^2}}{(\tilde{x}-1)+\sqrt{\eta^2(\tilde{y}+1)^2+(\tilde{x}-1)^2}}\right] +$$
$$(\tilde{x}-1) \ln\left[\frac{\eta(\tilde{y}-1)+\sqrt{\eta^2(\tilde{y}-1)^2+(\tilde{x}-1)^2}}{\eta(\tilde{y}+1)+\sqrt{\eta^2(\tilde{y}+1)^2+(\tilde{x}-1)^2}}\right] + \eta(\tilde{y}-1) \ln\left[\frac{(\tilde{x}-1)+\sqrt{\eta^2(\tilde{y}-1)^2+(\tilde{x}-1)^2}}{(\tilde{x}+1)+\sqrt{\eta^2(\tilde{y}-1)^2+(\tilde{x}+1)^2}}\right]$$
$$\tag{14b}$$

where $\tilde{x} = x/R$ and $\tilde{y} = y/R\eta$. Substituting Eq. (14a) into (9a) we finally obtain,

$$\frac{c^*(x,y)}{c_m} = 1 + \rho \frac{E_f}{E^*} \frac{R}{\pi h}\left[\Phi_m - \Phi\left(\frac{x}{R}, \frac{y}{R\eta}\right)\right] \tag{15}$$

with $\Phi_m = \int_0^1 \int_0^1 \Phi(\tilde{x}, \tilde{y}) d\tilde{x}\, d\tilde{y}$.

Figure 3 reports the solution of Eq. (15) substituted into (13) (solid line) and compared with that of Eq. (7) substituted into (13) (crosses and circles), for a square array. In this case we assumed $E_f = E$, $\nu = 0.5$, $R = 71.4\,a$ and $h = 5\,a$. For the numerical analysis, we considered the case of $d = 3.57\,a$ and $N = 1681$, giving $\rho = 0.25$, and that of $d = 7.14\,a$ and $N = 441$, with $\rho = 0.06$. Also in this case the accuracy increases as the fibril density $\rho$ increases.

We report in Figure 4 the force-versus-separation plot in dimensionless form for a circular and a square array with uniform compliance (solid lines) and with optimal compliance distribution, from Eq. (12) and (15), respectively, and same contact area (dashed lines). We adopt the same parameters used in the simulations in Figure 2 and 3 with highest $N$. The arrays having uniform compliance detach earlier and do not achieve ELS. Furthermore, the fibrils detach rapidly but not simultaneously once first detachment has occurred. The arrays with optimal compliance distribution instead obtain ELS reaching nearly maximum theoretical strength, with $F_{max} \approx Nf_{max}$. The displacements in the abscissa of Figure 4 are normalized by the nominal value $u_n = f_{max}/\pi a E^*$.

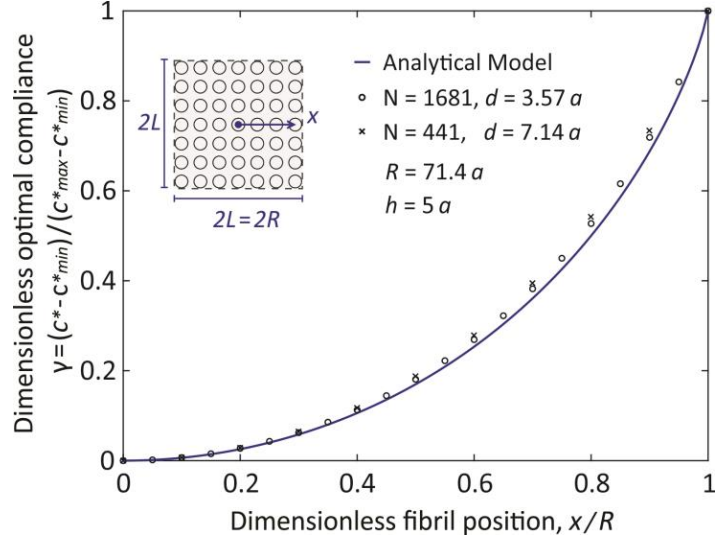

**Figure 3** Optimal compliance distribution for fibrils in a square array of size $2L$. Crosses and circles indicate the solution obtained with the numerical method, from Eq. (7), while the solid line reports the results from the proposed analytical model, from Eq. (15).

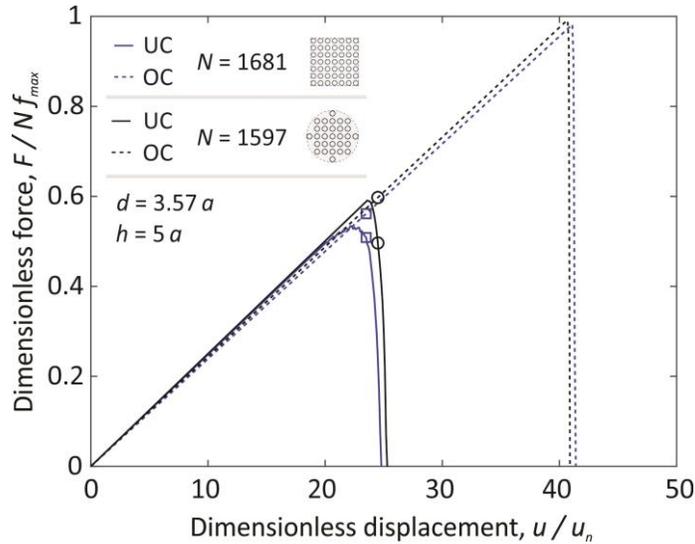

**Figure 4** Force versus displacement during detachment for a fibrillar adhesive having circular (black) and square (blue) array. We consider fibrils with uniform compliance (UC, solid lines) and with optimal compliance (OC, dashed lines). The hollow circle and square markers indicate the configuration shown in Figure 5 and 6, respectively.

Figures 5 and 6 show the distribution of fibril forces and the deformation of the fibrils and the backing layer for various cases, taken from the simulations in Figure 4 (circles and squares overlaid to the plots). Figure 5 shows the case of a circular array, while Figure 6 shows that of a square array. Both figures compare the case of uniform compliance distribution (left) with that of optimal compliance distribution (right). In this comparison, we keep the same separation $\bar{u}$ for uniform compliance and optimal compliance.

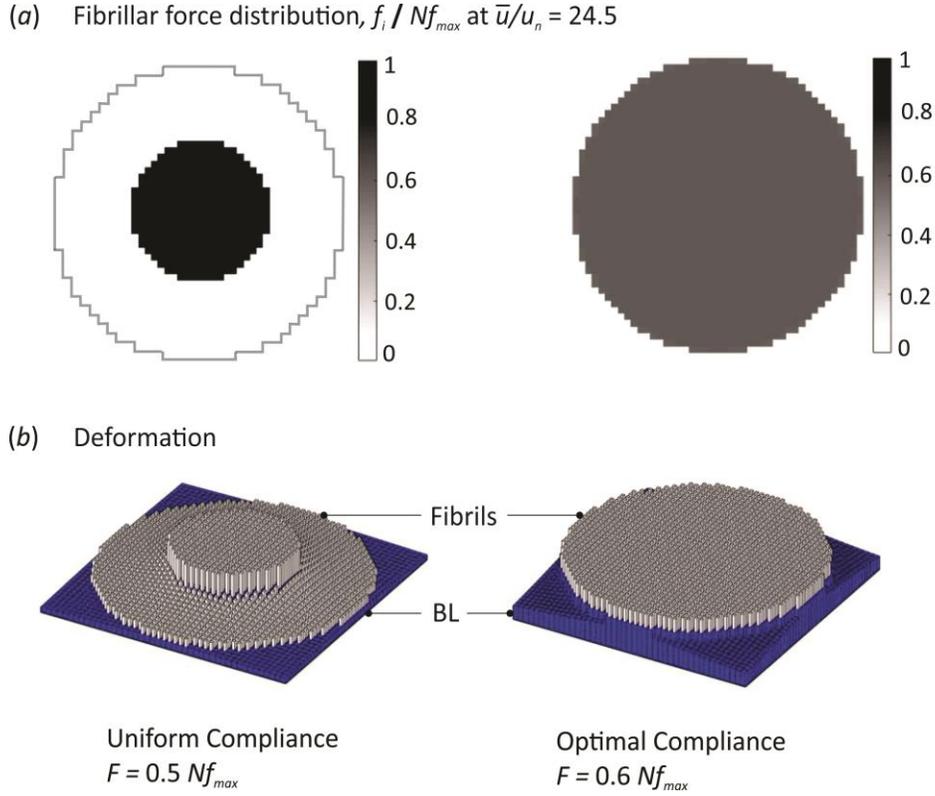

**Figure 5** Adhesive configuration, in terms of fibrillar force distribution (*a*) and BL and fibril deformation (*b*), for a circular array with uniform (left) and optimal (right) compliance distribution, from the analysis in Figure 4.

In Figure 7 we report the normalized detachment force (or adhesive strength) in the presence of misalignment for circular arrays (black lines) and square arrays (blue lines) optimized with the numerical solution from Eq. (7) (dashed lines) and with the proposed analytical solution from Eq. (9) (solid lines). In this figure we use the same parameters used in Figure 4 but different number of fibrils. The normalized detachment force, $F_{max}/Nf_{max}$, can be considered as a measure of the load sharing efficiency of the array. In our analysis, we only consider positive misalignment $\theta$, however, the results reported can be easily extended to negative misalignments in force of the symmetry of the problem given by $\theta_d = 0$. For both circular and square arrays, the numerical solution produces higher adhesive strength, compared to the analytical solution, for relatively small misalignments. For misalignments that are larger than a transition value $\theta_{tr}$, the analytical solution outperforms the numerical one, evidencing a benefit from the asymptotic approximation in the case of significant statistical misalignment (for $|\theta| > |\theta_{tr}|$). $\theta_{tr}$ is a function of the parameters defining the array, namely, number of fibrils $N$, fibril spacing $d$ and fibril averge length $h$. For the case analyzed in Figure 7, the transition misalignment is $\tan \theta_{tr} = 0.006$ ($\theta_{tr} = 0.006 \text{ rad} = 0.34°$), for the circular array, and $\tan(\theta_{tr}) = 0.021$ ($\theta_{tr} = 0.021 \text{ rad} = 1.23°$), for the square array. For both square and circular arrays, the adhesive strength produced with numerical optimization appears to be more sensitive to misalignment, compared to the strength produced by the proposed analytical solution.

(*a*) Fibrillar force distribution, $f_i / Nf_{max}$ at $\bar{u}/u_n = 23.5$

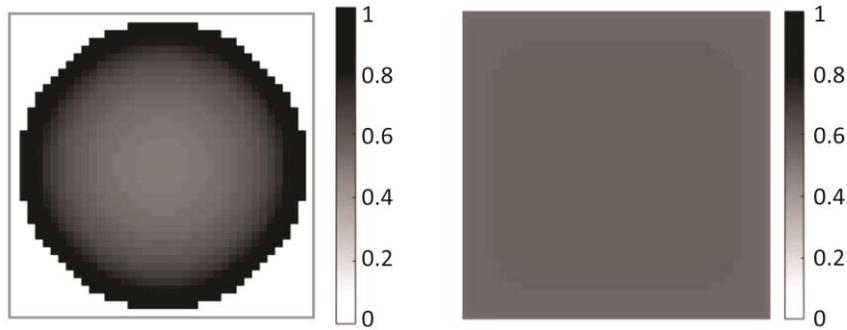

(*b*) Deformation

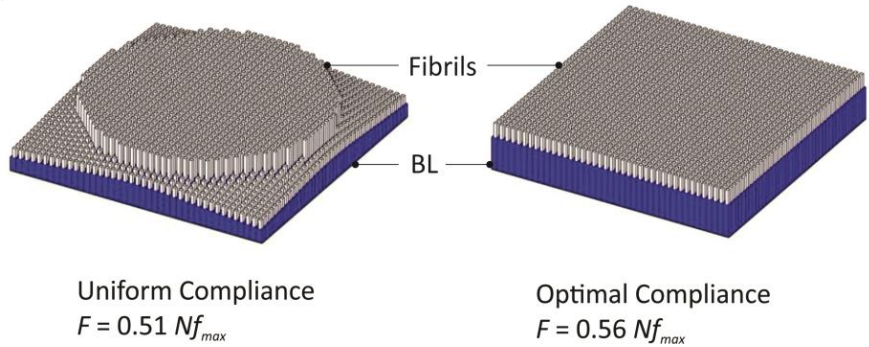

Uniform Compliance
$F = 0.51\ Nf_{max}$

Optimal Compliance
$F = 0.56\ Nf_{max}$

**Figure 6** Adhesive configuration, in terms of fibrillar force distribution (*a*) and BL and fibril deformation (*b*), for a square array with uniform (left) and optimal (right) compliance distribution, from the analysis in Figure 4.

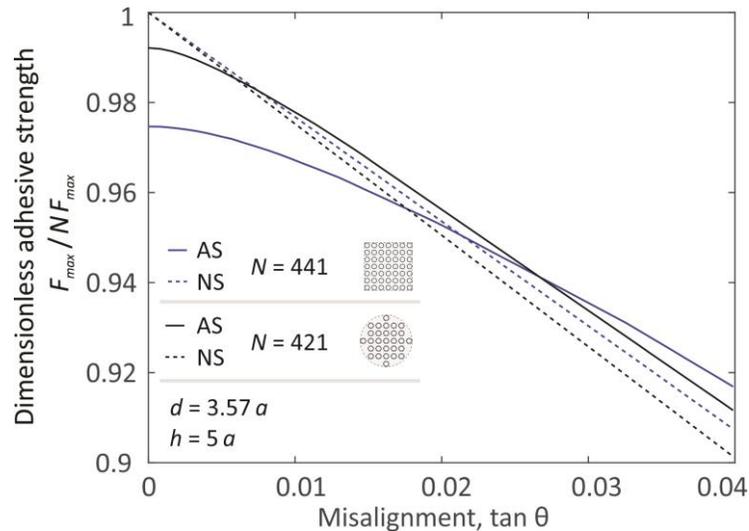

**Figure 7** Evolution of the dimensionless adhesive strength for increasing misalignment of arrays having optimal fibril compliance distribution. We consider square arrays (blue lines) and circular arrays (black lines) optimized with the numerical solution (NS, dashed lines), from Eq. (7), and with the proposed asymptotic solution (AS, solid lines), from Eq. (12) and (15).

**Discussion**

As reported in the *results* section, the adhesive strength obtained from arrays optimized with the proposed asymptotic solution show lower sensitivity to misalignment, compared to the ones optimized via numerical solution, and in some cases even higher value. The latter phenomenon is likely related to the significant stress redistribution generated by large misalignments, which challenges the benefits obtained from the optimal distribution of fibril compliance, since the latter was constructed under the hypothesis of $\theta_d = 0$. The benefit emerged from the asymptotic approximation suggests the possibility of exploring array optimization for a range of misalignment angles instead of simply assuming a specific value for $\theta_d$ when calculating the optimal compliance distribution. This, however, is beyond the scope of the present paper, hence is left for future work.

To achieve ELS at the array scale, functional grading of the fibril compliance distribution can potentially be done in multiple ways. For example, the Young modulus of the material composing the fibrils could be graded so that, for fibril $i$, we have $E_{f,i}/E_f = c_m/c_i^*$ substituted in Eq. (7), while for the continuum ensemble of fibrils we have $E_f(x,y)/E_f = c_m/c^*(x,y)$ substituted in Eq. (9). Another way to achieve optimal compliance distribution is tailoring the length of each fibril following the relation $h_i/h = c_i^*/c_m$ or $h(x,y)/h = c^*(x,y)/c_m$. In this case, different fibril lengths within the same array would make it difficult for all the fibrils to adhere perfectly to the RS. This is because the longest fibrils will likely undergo buckling, and therefore lose contact, in order to allow for the shortest ones to enter into contact. To counteract this effect, the RS or the BL should have a properly curved surface so that uniform contact across the interface can be achieved. The requirement for a specific curvature at the RS surface would significantly limit the applicability of the adhesive. The requirement of a properly curved BL surface, on the other hand, appears much less limiting, however the proposed model should be modified to account for this feature. Finally, another method to achieve ELS is functional grading of the stalk radius of the fibrils, following $a_i/a = \sqrt{c_m/c_i^*}$ or $a(x,y)/a = \sqrt{c_m/c^*(x,y)}$.

All the proposed methods above present numerous challenges from the manufacturing point of view, rendering it difficult to experimentally validate the proposed theory. An important design limitation related to functional grading of fibril compliance is the incremented risk of mutual adhesion among fibrils, also called fibril condensation [14], for the softest fibrils located at the perimeter of the contact region. This is because of the proportionality between axial stiffness and bending stiffness in a fibril, with bending stiffness being responsible for the prevention of fibril condensation. In this view, functional grading of the fibril modulus $E_{f,i}$ appears to be the most convenient strategy since both axial stiffness and bending stiffness are linearly proportional to it. The other two methods instead appear to be suboptimal since bending stiffness decreases more rapidly than axial stiffness for an increment of stalk length $h_i$ and a reduction of stalk radius $a_i$.

To mitigate the phenomenon of fibril condensation, one could reduce the contrast between maximum and minimum optimal compliance (in order to reduce the maximum compliance). Eq. (9) evidences how this contrast is proportional to the term $\rho \left(E_f/E^*\right)(R/h)$, suggesting its minimization as a viable strategy. At this purpose, one could reduce the contrast in optimal compliance by reducing the fibril stalk density $\rho$. From Eq. (7b) we can deduce that this would imply a reduction in the fibril stalk radius, hence a reduction in bending stiffness, therefore requiring proper consideration. A significant reduction in stalk radius would also increment the stress experienced by the fibril at detachment, incrementing the risk of failure of the material composing the fibril. Another approach is the reduction of the ratio $E_f/E^*$ by producing softer fibrils or stiffer BL. The reduction in fibril modulus comes again at the price of a reduction in bending stiffness but less so than a reduction in $\rho$. A stiffer BL would give instead more effective results but only in the presence of negligible misalignment. This is because a softer BL have been observed to better resist interfacial misalignment ($\theta \neq 0$) [13, 16]. Finally, one can reduce the ratio $R/h$ by incrementing the overall fibril length $h$ or by reducing the size of the array, via reduction of $R$. The former produces a significant loss in bending stiffness, as explained above, while the latter requires a reduction of the area of adhesive contact with consequent reduction of the maximum detachment force. This limitation can be mitigated by the division of the contact region into multiple sub-regions with independent arrays of fibrils. Such a hierarchical fibrillar subdivision in multiple arrays is often observed in nature but its development in engineering prototypes brings again new challenges.

**Conclusions**

The accuracy of our asymptotic model, in generating equal load sharing (ELS) for zero misalignment, increases with the density of the array. We performed multiple numerical simulations with various $N$ and $\rho$ for both circular and square arrays utilizing the asymptotic solution for optimal compliance distribution, from Eq. (12) and (15), respectively. For the circular arrays, the number of fibrils were varied from 208 to 1804 for a constant radius of $R/a = 80$. This varied the fibrillar density from 0.1 to 0.3 and resulted in a minimum load sharing efficiency of 0.98. For the square array, the number of fibrils was varied from 289 to 2401 for a constant radius of $R/a = 80$. This also varied the fibrillar density from 0.1 to 0.3, and again resulted in a minimum load sharing efficiency of 0.98. In conclusion, for both circular and square arrays, the results indicated a deviation in adhesive strength that is within 2%, compared to ELS. All this for $\theta_d = 0$ and $\theta = 0$, for square arrays and circular arrays, taken as representative cases for a wide range of practical applications. For the case of $\theta \neq 0$, the asymptotic approximation demonstrated some benefits in terms of reduced sensitivity of strength versus misalignment and even higher strength, for $|\theta| > |\theta_{tr}|$.

The proposed asymptotic analysis provides the numerous benefits of a closed-form solution, evidencing the determinants of an optimal interfacial stiffness distribution to reduce stress concentration, and hence achieve ELS, at the fibril array level. Our model system can be further generalized beyond fibrillar adhesives if used in an analogy that considers fibrils as brittle bonds uniformly distributed across an interface separated by a tensile load. Our solution indeed suggests a theoretical strategy to achieve maximum toughness of an adhesive interface by grading the stiffness of its bonds. Incremented strength of adhesive interfaces designed with proper compliance grading has been experimentally observed in shear adhesion [17]. Although shear adhesion is a different phenomenon than that analyzed in this paper, both adhesion mechanisms are controlled by stress concentration, hence the aforementioned experimental findings are to be taken as a qualitative validation of our theory.

**Acknowledgments**
The Natural Science and Engineering Research Council of Canada (NSERC) provided financial support, while the Institute for Computing, Information and Cognitive Systems (ICICS) provided logistic support.

## *Appendix A*

The condition of equal load sharing (ELS) is that at which each fibril tip transmit its maximum load at complete detachment, hence

$$f_i = f_{max}, \quad \forall i \tag{A1}$$

with all fibrils detaching simultaneously. This also implies that

$$u_i = \bar{u}_c + \tan\theta_d \, x_i, \quad \forall i \tag{A2}$$

with $\bar{u}_c$ the prescribed separation at detachment and $\theta_d$ the design misalignment. Substituting Eq. (A1) and (A2) in (3) we can obtain the expression

$$\frac{\bar{u}_c}{f_{max}} + \frac{\tan\theta_d}{f_{max}} x_i = \frac{16}{3\pi^2 E^* a_i} + c_i^* + \frac{1}{\pi E^*} \sum_{j \neq i} \frac{1}{r_{ij}}, \quad \forall i \tag{A3}$$

which, averaged over the whole array, gives

$$\frac{\bar{u}_c}{f_{max}} + \frac{\tan\theta_d}{f_{max}} \frac{1}{N}\sum_i x_i = \frac{16}{3\pi^2 E^*} \frac{1}{N}\sum_i \frac{1}{a_i} + c_m + \frac{1}{\pi E^*} \frac{1}{N}\sum_i \left(\sum_{j \neq i} \frac{1}{r_{ij}}\right) \tag{A4}$$

Equating now Eq. (A3) and (A4), we obtain

$$c_i^* = c_m + \frac{1}{\pi E^*}\left[\frac{1}{N}\sum_i \left(\frac{16}{3\pi}\frac{1}{a_i} + \sum_{j \neq i}\frac{1}{r_{ij}}\right) - \left(\frac{16}{3\pi}\frac{1}{a_i} + \sum_{j \neq i}\frac{1}{r_{ij}}\right)\right] + \frac{\tan\theta_d}{f_{max}}(x_i - x_m) \tag{A5}$$

Dividing now both sides of Eq. (A5) by $c_m$ and substituting from Eq. (6), (7b), (7c), (7d) and (7e), we finally obtain Eq. (7a).

## *Appendix B*

The critical separation between RS and BL in a general location of the interface identified by coordinates $(x, y)$ is

$$\bar{u}_c + \tan\theta_d \, x = u_{BL}(x,y) + \Delta h(x,y) \tag{B1}$$

with

$$\Delta h(x,y) = c^*(x,y) \, f_{max} \tag{B2}$$

the fibril elongation created by the transmission of the load $f_{max}$, and $\theta_d$ the design misalignment. Substitution of $\Delta h(x,y)$ from Eq. (B2) into (B1) gives,

$$\frac{\bar{u}_c}{f_{max}} + \frac{\tan\theta_d}{f_{max}}x = \frac{u_{BL}(x,y)}{f_{max}} + c^*(x,y) \tag{B3}$$

which averaged over the whole contact region gives,

$$\frac{\bar{u}_c}{f_{max}} + \frac{\tan\theta_d}{f_{max}}x_m = \frac{u_{BL,m}}{f_{max}} + c_m \tag{B4}$$

with $x_m = \frac{1}{A}\int_A x\, dA$, the $x$ coordinate of the centroid of the contact region, and $u_{BL,m} = \frac{1}{A}\int_A u_{BL}(x,y)dA$, the average BL displacement. Substitute now $u_{BL}(x,y)$ from (9b), and the same for $u_{BL,m}$. Equating then Eq. (B3) and (B4), we obtain

$$c^*(x,y) = c_m + \frac{R}{E^* d^2}[\tilde{u}_{BL,m} - \tilde{u}_{BL}(x,y)] + \frac{\tan\theta_d}{f_{max}}(x - x_m) \tag{B5}$$

Finally, dividing both sides of Eq. (B5) and substituting from Eq. (6), (7b), and (7e) we have Eq. (9a).

### *Appendix C*

A general formulation for the displacement of an elastic half-space subject to the uniform pressure $f_{max}/d^2$ over the contact region of area $A$, obtained from linear elastic contact mechanics [18], can be written as

$$u_{BL}(x,y) = \frac{f_{max}}{\pi E^* d^2}\int_A \frac{dX\,dY}{\sqrt{(X-x)^2+(Y-y)^2}} \tag{C1}$$

Substituting Eq. (C1) into Eq. (9b) we obtain

$$\tilde{u}_{BL}(x,y) = \frac{1}{\pi R^2}\int_A \frac{R\,dX\,dY}{\sqrt{(X-x)^2+(Y-y)^2}} \tag{C2}$$

Let us now divide the contact region into $N$ sub-regions of area $d^2$, with $N \to \infty$. We can assume then $dX\,dY = A/N = d^2$, hence we can rewrite Eq. (C2) for $x = x_i$ and $y = y_i$, with the use of Eq. (7d), as

$$\tilde{u}_{BL}(x_i, y_i) = \lim_{N \to \infty} \frac{\Gamma}{N}\sum_{j \neq i}\frac{R}{\sqrt{(x_j-x_i)^2+(y_j-y_i)^2}} \tag{C3}$$

If we now consider small variation in stalk radius among fibrils ($a_i \simeq a$) we can neglect the first term in Eq. (7c), since it is constant among fibrils and therefore redundant in the difference $\psi_m - \psi_i$ in Eq. (7a), giving in this case

$$\psi_i = \frac{\Gamma}{N}\sum_{j \neq i}\frac{R}{r_{ij}} \tag{C4}$$

By substituting $r_{ij}^2 = (x_j - x_i)^2 + (y_j - y_i)^2$ into Eq. (C3) and the result into Eq. (10), we finally obtain Eq. (C4).